\documentclass{article}

\usepackage{arxiv}

\usepackage[utf8]{inputenc} \usepackage[T1]{fontenc}    

\PassOptionsToPackage{hyphens}{url}

\usepackage[breaklinks,backref=page]{hyperref}

\usepackage{url}            \usepackage{booktabs}       \usepackage{amsmath}        \usepackage{amsfonts}       \usepackage{nicefrac}       \usepackage{microtype}      \usepackage{lipsum}         \usepackage{graphicx}
\usepackage[numbers,sort&compress]{natbib}
\usepackage{doi}
\usepackage{float}
\usepackage{multirow}
\usepackage{lineno}

\usepackage{cleveref}

\usepackage{wrapfig}

\usepackage[justification=centering]{caption}

\usepackage[inline]{enumitem}

\usepackage{bbding}

\usepackage[flushleft]{threeparttable}
\usepackage{tabularx}
\usepackage{array}

\usepackage{listings}

\usepackage{subcaption}

\usepackage{orcidlink}

\usepackage{pgffor}

\usepackage{siunitx}
\renewcommand*{\backref}[1]{}
\renewcommand*{\backrefalt}[4]{
  \ifcase #1
    No citations.\or
(Cited on page: #4).
  \else
(Cited on pages: #4).
  \fi
}

\title{Esports and Physiological Tremor a StarCraft~2 Tournament Study
 }

\date{}

\author{
  \textbf{Andrzej Białecki}\thanks{Corresponding author: \texttt{andrzej.bialecki94@gmail.com}}\,,\vspace{0.1cm}
  \textbf{Izabela Ghafour},\vspace{0.1cm}
  \textbf{Szymon Kuliś},\vspace{0.1cm}
  \textbf{Maciej Skorulski},\vspace{0.1cm}
  \textbf{Jan Gajewski}
}

\hypersetup{
pdftitle={Esports and Physiological Tremor a StarCraft 2 Tournament Study},
pdfauthor={Andrzej Białecki, Izabela Ghafour, Szymon Kuliś, Maciej Skorulski, Jan Gajewski},
pdfkeywords={esports, tremor, tournament, biomechanics, StarCraft 2},
}

\providecommand{\parencite}[1]{}

\renewcommand{\parencite}[1]{\cite{#1}}

\begin{document}
\renewcommand{\figureautorefname}{Fig.}
\renewcommand{\subsectionautorefname}{Subsection}

\maketitle

\begin{abstract}

  Physiological tremor of the upper limb is a sensitive neuromuscular indicator that may be modulated by cognitive load and competitive stress, yet its behaviour in real esports conditions remains uncharacterised. We measured wrist accelerometer-based tremor in 16 healthy adult male StarCraft~2 players across two tournament days, computing log power spectral density ($log(PSD)$) and dominant frequency in four bands (2--4, 8--14, 10--20, and 1--25Hz) and comparing them to published population norms using linear mixed models. Players deviated significantly from the reference in all bands: $log(PSD)$ was elevated at 2--4~Hz and substantially reduced at higher frequencies (Cohen's $d = 1.6$--$2.3$), suggesting long-term neuromuscular adaptation to the fine-motor demands of esports. Tremor indicators declined systematically over the tournament day. Contrary to the fatigue-related increases typical of traditional motor tasks. Neither game outcome nor actions per minute significantly predicted post-game tremor. These findings suggest physiological tremor may reflect a generalised psychophysiological adaptation to competitive esports rather than being a short-term performance predictor.

\end{abstract}


\section{Introduction}
\label{sec:introduction}

Recent advancements in technology have enabled the growth of new media such as gaming and electronic sports (esports)~\cite{Cramer2021}. This new venue of human activity opens a world in which everyone is connected via high-speed internet~\cite {Chap2023}. Although there are many definitions of esports~\cite{Freeman2017}, and their recognition as a sports category is not universal~\cite {Parry2019}, this area of human interest has been classified as high-performance human-computer interaction~\cite{Watson2021}. For the sake of our article, we assume that ``Esports is a group of sports that use electronics and telecommunication technologies to create the means of competition for their users that intend to maximize their skill''~\cite{Bialecki2022Redefining}. While there are many video games to choose from, there is no doubt that StarCraft~2 is one of the long-standing esports titles~\cite{Scholz2011,Rea2016CraftingStars}. Even in the absence of the aforementioned global recognition, the academic interest in this field is evident in the growing number of articles published on the topic~\cite{Poulus2026,Bialecki2024Scoping}.

Research on esports is especially rewarding due to the apparent lack of knowledge on the training methodologies~\cite{Bialecki2024Scoping}. Similar to traditional sports, all areas of research on human activity and performance optimization could be applied in esports, e.g., biomechanics~\cite{Dupuy2024}, psychology~\cite{Tholl2024,DiFranciscoDonoghue2021,Poulus2026}, and physiology~\cite{Sadowska2023}, among others. Gaming, and therefore esports, is a multidisciplinary area of research; fully understanding the issues faced by many stakeholders requires knowledge of many adjacent areas~\cite{Scholz2019Stakeholders}. Furthermore, the issue of moving forward with practical esports and gaming research stems from the many game genres in existence~\cite{Bialecki2024Scoping}. Each type differs in rules, display methods, and required skills. The fragmented nature of games introduces the need for specialization; understanding the differences between games and the motivations for play is crucial~\cite{Jang2020,Bickmann2021,Ma2021}. While some works exist to differentiate between game genres using biomechanical indicators~\cite{Dupuy2025}, it is not entirely clear how the tournament load affects players' performance in these areas. Hotkey use could contribute to the overall biomechanical load sustained during such events. Interestingly, previous research did show that hotkey use is one of the key predictors, along with the number of perception-action cycles (PACs)~\cite{Thompson2013}. In the end, objective physiological markers of load in esports remain limited~\cite{Wu2025}, and we ought to aim to quantify the tournament load players need to withstand.

Physiological tremor refers to an inherent, involuntary oscillatory motion observed in every individual~\cite{Kosmowska2021,Shanghavi2024}. It arises from the interaction between central neural drive, peripheral feedback, and mechanical properties of the musculoskeletal system~\cite{Son2019}. This benign phenomenon, often imperceptible to the naked eye, does not typically impede motor function in healthy individuals~\cite{Namburi2025}. Physiological tremor is considered a sensitive indicator of neuromuscular function~\cite{White2019}. Our research direction is further motivated by the perceived physical exertion being a byproduct of gaming or esports activities~\cite{Tholl2025}.

Competitive gaming involves prolonged static postures, repetitive movements, and high cognitive stress~\cite{Palanichamy2020}. These conditions are known to elevate physiological stress, potentially intensifying tremor amplitude and adversely affecting the fine motor control required for precise in-game actions~\cite{Nicholson2024}. Enhanced physiological tremor is observed when anxiety is present~\cite{Lenka2021}. Therefore, investigating tremor during esports tournaments may provide novel insights into player load and the psychophysiological responses associated with high-pressure competitive environments.

Data indicate that gaming sessions lasting several hours affect hemodynamic and metabolic parameters, such as pulse wave velocity (PWV) and energy expenditure (EE). Prolonged exposure to gaming poses a real challenge to the body's circulatory and metabolic systems~\cite{Ketelhut2024}. Competing for financial rewards, prestige, and ranking may therefore trigger stress responses in a manner functionally similar to that observed in traditional sports~\cite{LeisPhD2022}.

Our understanding of physiological tremor in real tournament conditions remains limited. A clear need for research in conditions as close as possible to real competition, rather than solely in laboratory settings, is clear to us. We hope to explore and provide a better understanding of how the combination of high cognitive load, pressure to perform, and somatic stress symptoms affects the precision of upper-limb movements in esports athletes. Given available theoretical evidence and the practical needs of esports teams regarding the prevention of overexertion, performance decline, and burnout, this study aimed to investigate upper-limb physiological tremor during a StarCraft~2 tournament. The analysis was designed to verify the hypothesis that very high cognitive load, resulting from the need to perform hundreds of actions per minute, combined with the pressure of the tournament environment, significantly modulates the characteristics of a player's muscle tremor.

Our goal for this research was to address a gap in the literature and investigate upper-limb physiological tremor during a StarCraft~2 tournament, using it as a metric leveraged in other sports, as indicated above. To reach our goal, we have formulated the following research questions:
\begin{enumerate*}[label=\textbf{RQ\arabic*:}, itemjoin={\quad}, itemjoin*={\quad}]
  \item Does the tournament load cause a significant deviation of the players' tremor from the population norm?
  \item Does game outcome or otherwise player performance predict post-game tremor levels?
\end{enumerate*}

\section{Material and Methods}
\label{sec:material_and_methods}

\subsection{Participants}

Data for this study were collected in collaboration between the research team and the main tournament organizer. During two tournaments from a series: ``H.4.0.S Galactic Battle''. The tournament names were as follows:
\begin{enumerate*}[label=(\arabic*)]
  \item ``H.4.0.S~Galactic~Battle~3~by~ENDORFY'',
  \item ``H.4.0.S~Galactic~Battle~Champions''.
\end{enumerate*}
The distribution of qualified and invited players, as well as the players' main selected in-game races, is shown in~\autoref{tab:tournament_races}. All study participants (n=16) were healthy adult males aged 20-29.

\begin{table*}[h]
  \centering
  \caption{Player Main In-game Race Distribution, Selection, and Tremor Measurement Across Tournaments}
  \label{tab:tournament_races}
  \begin{tabular}{lccccccc}
    \toprule
                                        & \multicolumn{2}{c}{Selection} & \multicolumn{3}{c}{Race} &         &                                    \\
    \cmidrule(lr){2-3} \cmidrule(lr){4-6}
    Tournament                          & Invited                       & Qualified                & Protoss & Terran & Zerg & Total  & Measured  \\
    \midrule
    H.4.0.S Galactic Battle 3 (ENDORFY) & 7                             & 9                        & 9       & 5      & 2    & 16     & 15        \\
    H.4.0.S Galactic Battle Champions   & 0                             & 8                        & 5       & 1      & 2    & 8      & 8         \\
    \midrule
    Total                               & 7                             & 17                       & 14      & 6      & 4    & 18$^*$ & 16$^{**}$ \\
    \bottomrule
  \end{tabular}
  \par\smallskip
  \footnotesize\textit{Note.} $^*N = 18$ unique players across both tournaments; some players competed in both. Race assigned by majority of games played within each tournament. $^{**}N = 16$ unique players with valid tremor measurements.
\end{table*}

Each of the 16 players contributed between 4 and 38 valid measurements ($18.8 \pm 13.0$; $N_{\text{total}} = 300$).

\subsection{Methods}

Tremor measurement involved participants being seated with their back supported and their elbows flexed at approximately 90 degrees. Furthermore, the forearm was held in a ``neutral'' position. To assure the ability to compare our results to the previously established reference values, we have followed the same measurement protocol as defined in~\cite{Gajewski2023}, namely: ``The accelerometer was positioned over the participant's wrist. The 1 kg inertial load~\cite{Tomczak2014} did not exceed 10\% of the elbow flexors' MVC (maximum voluntary contraction), even in potentially weaker participants. Therefore, it was assumed that the external load applied did not cause fatigue that could affect physiological tremor.''. The acceleration was recorded using the ZPP-3D/BC Acceleration Measurement Set (JBA Zb. Staniak, Poland).

\begin{figure*}[ht]
  \centering
  \includegraphics[width=\textwidth]{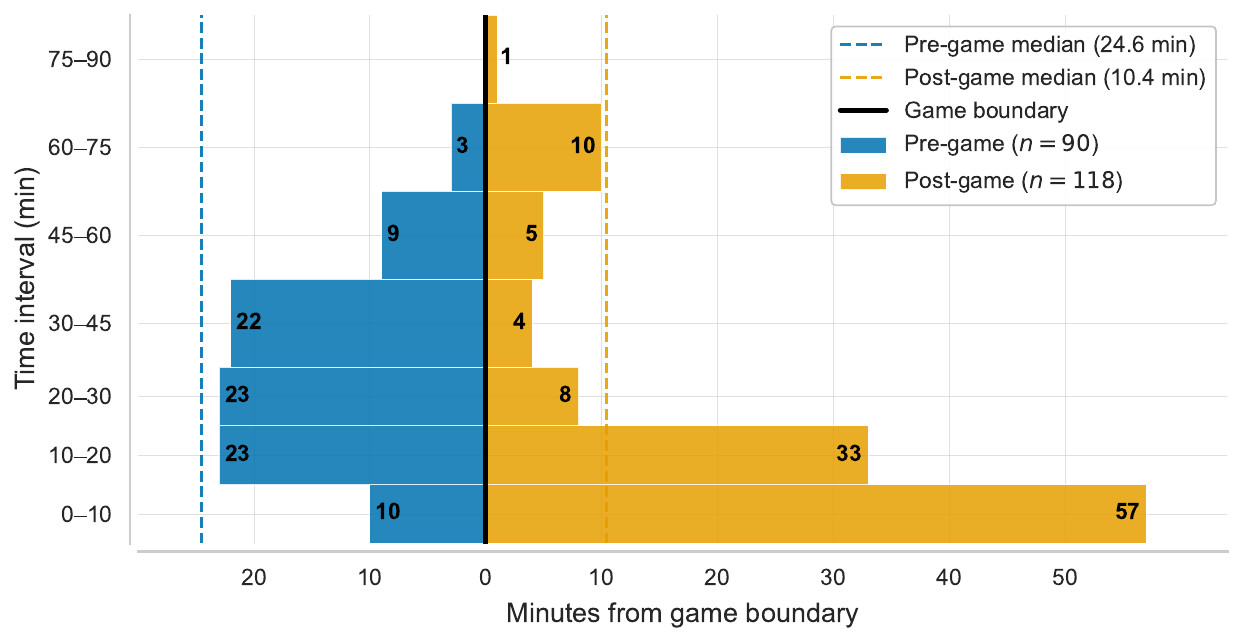}
  \caption{Distribution of tremor measurement timing relative to matched games. Pre-game measurements (left) reflect the interval between measurement and game start; post-game measurements (right) reflect the interval between game end and measurement. Dashed lines indicate medians.}
  \label{fig:timing_butterfly}
\end{figure*}

The measurement of tremor was collected as soon as possible after the end of a match. However, due to the tournament environment, the data had to be collected with varying timing between games. The timing of tremor measurements relative to matched games is visualized in~\autoref{fig:timing_butterfly}.

A standard signal of forearm tremor has two characteristic peaks at 3 Hz and around \SIrange{10}{12}{\hertz}. Most of the signal (98\%) is in the range of \SIrange{1}{25}{\hertz}. The low-frequency band consists of frequencies in the \SIrange{2}{4}{\hertz} range. The high frequency range is defined as \SIrange{8}{14}{\hertz} and the band most sensitive to fatigue is \SIrange{10}{20}{\hertz}. As the tremor amplitude depends on several factors, other indicators should be used in assessment. The power spectral density (PSD) of the frequency signal is comparable across individuals, so selected variables for the analysis should rely on it. We chose a Log amplitude indicator defined by~\autoref{eq:log_amplitude_indicator}~\cite{Gajewski2023}:
\begin{equation}
  \label{eq:log_amplitude_indicator}
  LAI(f_1, f_2)=\frac{1}{f_2-f_1}\int^{f_2}_{f_1}lnPSD(f)df
\end{equation}
and a mean frequency of power components as in~\autoref{eq:mean_frequency} \cite{Gajewski2023}:
\begin{equation}
  \label{eq:mean_frequency}
  F(f_1,f_2)=\frac{\int^{f_2}_{f_1}f \cdot PSD(f)df}{\int^{f_2}_{f_1}PSD(f)df}
\end{equation}
where the interpretation of the notation is as follows:
\begin{itemize}
  \item $LAI(f_1,f_2)$ -- log amplitude indicator of frequency band between $f_1$ and $f_2$,
  \item $F(f_1,f_2)$ -- mean frequency power,
  \item $PSD(f)$ -- power spectral density function,
  \item $f_1, f_2$ -- the boundaries of the frequency band.
\end{itemize}
The calculation of indicators was performed across the four listed frequency bands, and results from \cite{Gajewski2023} were used as a reference. The cited study collected data from 276 healthy men practicing various sports, such as canoeing, swimming, volleyball, wrestling, baseball, ice hockey, archery, and others.

Code required for processing the data was written using Python version ``3.13.0''~\cite{Python}. Data analyses were performed using Pingouin~\cite{Vallat2018Pingouin}, statsmodels~\cite{Seabold2010statsmodels}, numpy~\cite{Harris2020numpy}, and pandas~\cite{mckinney2010pandas,Reback2020pandas}. Finally, the plots were created using matplotlib~\cite{Hunter2007Matplotlib}, and seaborn~\cite{Waskom2021}. The main statistical methods were linear mixed models (LMM), one-sample $t$-tests, repeated-measures analysis of variance (ANOVA), and Spearman rank correlation. Ordinary least squares~(OLS) regression was also used to overlay per-player and pooled trend lines in figures for visual interpretation; OLS estimates were not used for inference.

\section{Results}
\label{sec:results}

\subsection{Tremor Deviation from Reference}

\begin{figure*}[ht]
  \centering
  \includegraphics[width=\linewidth]{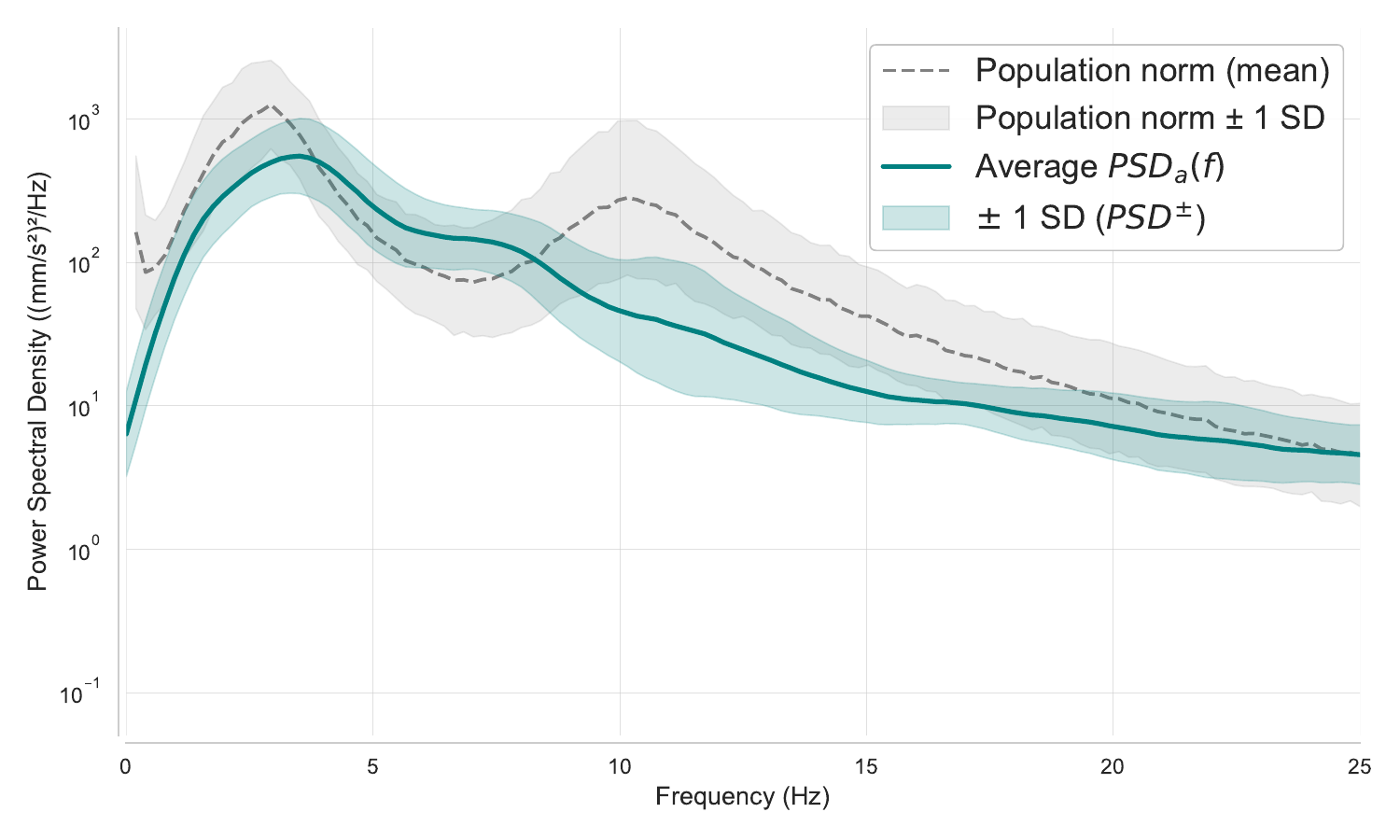}
  \caption{Average Power Spectral Density (PSD) for the measured group.}
  \label{fig:average_psd}
\end{figure*}

Our initial task was to verify the data's utility. Group-averaged Power Spectral Density ($log(PSD)$) is shown in \autoref{fig:average_psd}. Manual verification of the recordings revealed that the data quality was sufficient to proceed and that there were no major artifacts that could affect the results.

To test whether players deviated from the reference, linear mixed-model intercepts were estimated per band, accounting for repeated measurements per player; the results of this method are available in \autoref{tab:mixed_model}. Players deviated significantly from the reference in all bands for $log(PSD)$ and mean frequency. For $log(PSD)$, the \SIrange{2}{4}{\hertz} band showed a significant positive deviation, while all higher bands showed significant negative deviations. For mean frequency, the pattern was similar except for the \SIrange{10}{20}{\hertz} band, which showed a significant positive deviation ($M = 1.83$). Effect sizes were large for the \SIrange{8}{14}{\hertz}, \SIrange{10}{20}{\hertz}, and \SIrange{1}{25}{\hertz} bands (Cohen's $d = 1.6$--$2.3$ for $log(PSD)$), indicating that esports players under competitive conditions show substantially lower tremor power at physiologically relevant frequencies compared to the ``regular'' athletes.
As a secondary analysis, a repeated-measures ANOVA confirmed that the magnitude of deviation differed significantly across frequency bands for both $log(PSD)$ ($F = 132.92$, $\eta^2_p = 0.58$, $p < .001$) and Mean Frequency ($F = 63.36$, $\eta^2_p = 0.78$, $p < .001$), indicating that bands were not uniformly affected. Confirmatory one-sample $t$-tests were found to be significant across the board with $p < .001$ for all metrics.

\begin{table}[htbp]
  \centering
  \small
  \setlength{\tabcolsep}{4pt}
  \renewcommand{\arraystretch}{1.08}
  \caption{Linear Mixed-Model Intercepts: Player Deviations from Population Norm}
  \label{tab:mixed_model}
  \begin{threeparttable}
    \begin{tabularx}{\columnwidth}{@{}l c c c >{\raggedright\arraybackslash}p{0.30\columnwidth} >{\centering\arraybackslash}p{0.10\columnwidth}@{}}
      \toprule
      \multicolumn{6}{l}{\textit{Log PSD}}                                                                                                                                  \\
      Band      & $n$ & $N$ & $M$               & 95\% CI                               & \textit{p}                                                                        \\
      \midrule
      2--4~Hz   & 16  & 300 & $\phantom{-}0.55$ & $[\phantom{-}0.14,\ \phantom{-}0.96]$ & $.009$$^{**}$ \\
                                                                                                        8--14~Hz & 16 & 300 & $-1.49$ & $[-1.88,\ -1.10]$ & $.001$$^{***}$  \\
      10--20~Hz & 16  & 300 & $-1.51$           & $[-1.83,\ -1.18]$                     & $.001$$^{***}$ \\
                                                                                                         1--25~Hz & 16 & 300 & $-1.12$ & $[-1.47,\ -0.77]$ & $.001$$^{***}$ \\
      \midrule
      \multicolumn{6}{l}{\textit{Mean Frequency}}                                                                                                                           \\
      Band      & $n$ & $N$ & $M$               & 95\% CI                               & \textit{p}                                                                        \\
      \midrule
      2--4~Hz   & 16  & 300 & $\phantom{-}1.06$ & $[\phantom{-}0.85,\ \phantom{-}1.27]$ & $.001$$^{***}$ \\
                                                                                                         8--14~Hz & 16 & 300 & $-0.82$ & $[-1.11,\ -0.53]$ & $.001$$^{***}$ \\
      10--20~Hz & 16  & 300 & $\phantom{-}1.83$ & $[\phantom{-}1.40,\ \phantom{-}2.26]$ & $.001$$^{***}$ \\
                                                                                                         1--25~Hz & 16 & 300 & $-0.49$ & $[-0.71,\ -0.28]$ & $.001$$^{***}$ \\
      \bottomrule
    \end{tabularx}
    \begin{tablenotes}[flushleft]
      \small
      \item \textit{Note.} $^{*}p < .05$. $^{**}p < .01$. $^{***}p < .001$.
    \end{tablenotes}
  \end{threeparttable}
\end{table}

\subsection{Temporal Drift of Tremor Indicators}

To test whether tremor indicators changed systematically over the course of the tournament day, we fitted linear mixed models of the form ``$\text{value} \sim \text{time\_elapsed\_min} + (1 \mid \text{player})$'', with time elapsed since each player's first measurement on that day as the predictor and a random intercept per player, fitted by restricted maximum likelihood (REML). Significance of the slope was assessed with a Wald $z$-test on the fixed-effect coefficient. Results are shown in \autoref{tab:temporal_lmm}.

\begin{table*}[htbp]
  \centering
  \scriptsize
  \setlength{\tabcolsep}{3pt}
  \renewcommand{\arraystretch}{1.05}
  \caption{Linear mixed model estimates for temporal drift in standardised tremor indicators across the tournament day. $\beta$ is the change in standardised units per minute of elapsed time, estimated from \texttt{value\,\textasciitilde\,time\_elapsed\_min + (1\,|\,player)}.}
  \label{tab:temporal_lmm}
  \begin{tabular*}{\linewidth}{@{\extracolsep{\fill}}l S[table-format=2.0] S[table-format=3.0] r >{\raggedright\arraybackslash}p{0.31\linewidth} >{\centering\arraybackslash}p{0.11\linewidth}@{}}
    \toprule
    Band & {$n_{\text{players}}$} & {$N$} & {$\beta$} & 95\% CI & {$p$} \\
    \midrule
    \multicolumn{6}{l}{\textit{Log PSD}} \\
    2-4 & 16 & 300 & $-0.0004$ & $[-0.0009,\ \phantom{-}0.0002]$ & $.201$ \\
    8-14 & 16 & 300 & $-0.0010***$ & $[-0.0015,\ -0.0004]$ & $<.001$ \\
    10-20 & 16 & 300 & $-0.0012***$ & $[-0.0017,\ -0.0006]$ & $<.001$ \\
    1-25 & 16 & 300 & $-0.0010***$ & $[-0.0016,\ -0.0004]$ & $<.001$ \\
    \midrule
    \multicolumn{6}{l}{\textit{Mean Frequency}} \\
    2-4 & 16 & 300 & $-0.0003$ & $[-0.0007,\ \phantom{-}0.0002]$ & $.221$ \\
    8-14 & 16 & 300 & $-0.0006*$ & $[-0.0011,\ \phantom{-}0.0000]$ & $.035$ \\
    10-20 & 16 & 300 & $\phantom{-}0.0000$ & $[-0.0005,\ \phantom{-}0.0005]$ & $.980$ \\
    1-25 & 16 & 300 & $-0.0005**$ & $[-0.0009,\ -0.0001]$ & $.008$ \\
    \bottomrule
  \end{tabular*}
  \vspace{2pt}

  \footnotesize{*$p < .05$, **$p < .01$, ***$p < .001$.}
\end{table*}

In the temporal analysis figure, each coloured point is one measurement; thin coloured lines show per-player ordinary least squares (OLS) trends and the thick line(s) show the pooled OLS trend. The annotated $\beta$ values and significance stars are taken directly from the linear mixed model reported in the accompanying table. Standardised values are z-scores relative to population norms. The $log(PSD)$ indicator declined significantly over the tournament day in the \SIrange{8}{14}{\hertz}, \SIrange{10}{20}{\hertz}, and \SIrange{1}{25}{\hertz} bands ($\beta \approx -0.001$ SD/min, $p < .001$). Over a full tournament day of approximately 300 minutes the cumulative drift was modest in magnitude but consistent across players with a value of roughly $0.3$~SD. The \SIrange{2}{4}{\hertz} band showed no significant temporal trend.

\begin{figure*}[htbp]
  \centering
  \includegraphics[width=\linewidth]{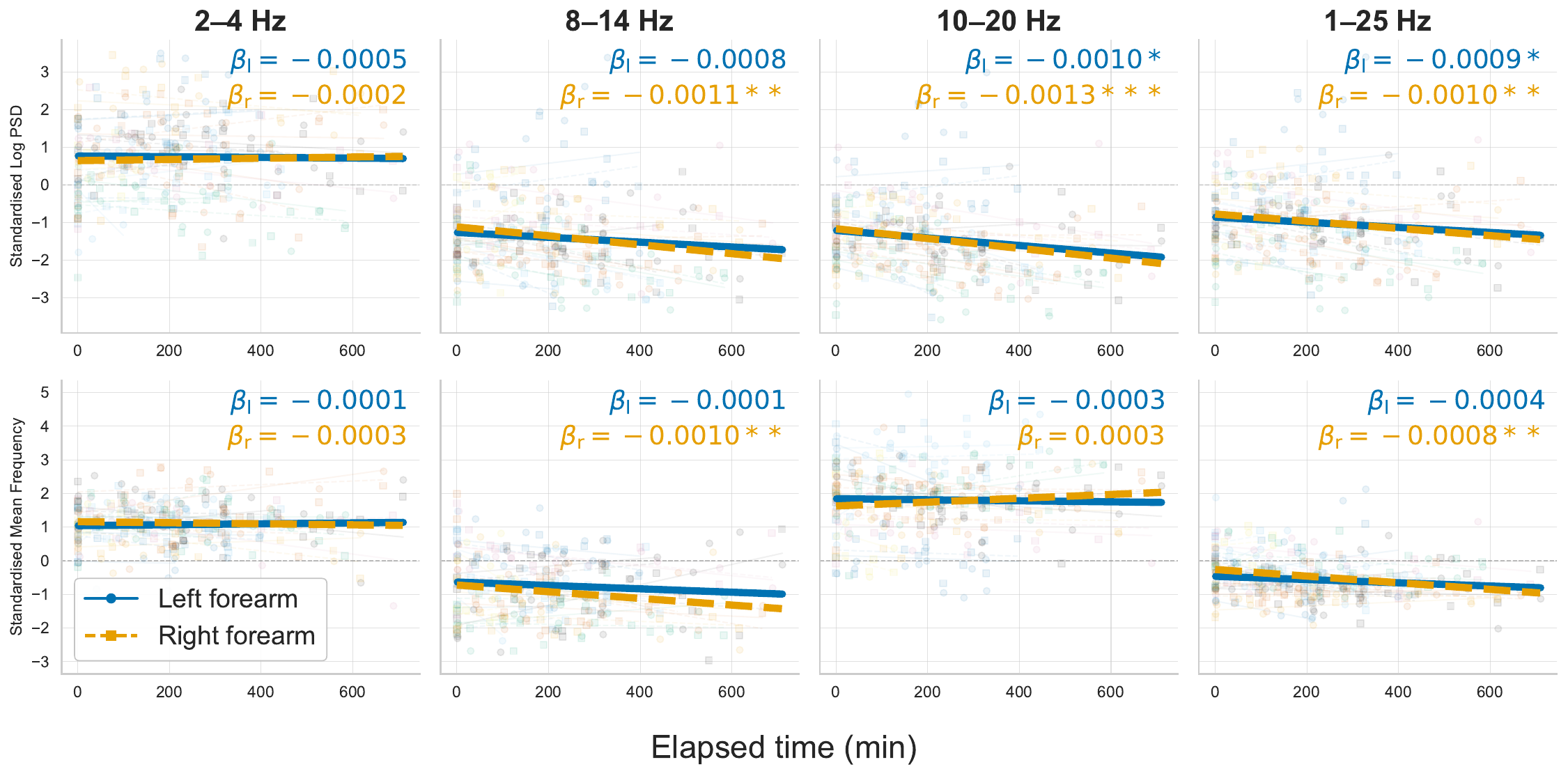}
  \caption{Temporal drift of forearm tremor indicators (by side) over the tournament day.}
  \label{fig:temporal_trends_by_hand}
\end{figure*}

To test whether the two forearms drifted at different rates, the same modelling approach (REML-fitted linear mixed models with a random intercept per player, significance assessed via Wald $z$-tests) was applied separately to left and right-forearm measurements, alongside a combined model including a time $\times$ forearm interaction term. Results are shown in \autoref{tab:temporal_lmm_by_hand} and \autoref{fig:temporal_trends_by_hand}, where the pooled trends are split by forearm (blue solid = left, red dashed = right). Both forearms showed a similar negative drift in $log(PSD)$ for the \SIrange{10}{20}{\hertz} and \SIrange{1}{25}{\hertz} bands, and the right forearm additionally drifted in \SIrange{8}{14}{\hertz} $log(PSD)$ and Mean Frequency ($p < .01$); however, the time $\times$ forearm interaction was not significant for any band or indicator (all $p \geq .125$), so the rate of drift did not differ reliably between forearms.

\begin{table}[htbp]
  \centering
  \small
  \setlength{\tabcolsep}{4pt}
  \renewcommand{\arraystretch}{1.08}
  \caption{Linear mixed model slopes ($\beta$, standardised units per minute) for temporal drift in tremor indicators, fit separately for the left and right forearm, from \texttt{value\,\textasciitilde\,time\,*\,forearm + (1\,|\,player)}. Asterisks on $\beta$ indicate whether that forearm's slope differs significantly from zero (i.e., a significant drift over time). The right-most $p$-value is from the time$\times$forearm interaction term and tests whether the rate of change differs between forearms, independent of the per-forearm significance of $\beta$.}
  \label{tab:temporal_lmm_by_hand}
  \begin{threeparttable}
    \begin{tabularx}{\columnwidth}{@{}>{\raggedright\arraybackslash}p{0.10\columnwidth} r >{\centering\arraybackslash}X r >{\centering\arraybackslash}X >{\centering\arraybackslash}p{0.12\columnwidth}@{}}
      \toprule
            & \multicolumn{2}{c}{Left forearm} & \multicolumn{2}{c}{Right forearm} &                                    \\
      \cmidrule(lr){2-3} \cmidrule(lr){4-5}
      Band  & $N$                              & $\beta$                           & $N$ & $\beta$             & $p$    \\
      \midrule
      \multicolumn{6}{l}{\textit{Log PSD}}                                                                              \\
      2-4   & 153                              & $-0.0005$                         & 147 & $-0.0002$           & $.600$ \\
      8-14  & 153                              & $-0.0008$                         & 147 & $-0.0011**$         & $.310$ \\
      10-20 & 153                              & $-0.0010*$                        & 147 & $-0.0013***$        & $.570$ \\
      1-25  & 153                              & $-0.0009*$                        & 147 & $-0.0010**$         & $.616$ \\
      \midrule
      \multicolumn{6}{l}{\textit{Mean Frequency}}                                                                       \\
      2-4   & 153                              & $-0.0001$                         & 147 & $-0.0003$           & $.479$ \\
      8-14  & 153                              & $-0.0001$                         & 147 & $-0.0010**$         & $.352$ \\
      10-20 & 153                              & $-0.0003$                         & 147 & $\phantom{-}0.0003$ & $.125$ \\
      1-25  & 153                              & $-0.0004$                         & 147 & $-0.0008**$         & $.175$ \\
      \bottomrule
    \end{tabularx}
    \begin{tablenotes}[flushleft]
      \small
      \item \textit{Note.} $^{*}p < .05$. $^{**}p < .01$. $^{***}p < .001$.
    \end{tablenotes}
  \end{threeparttable}
\end{table}

\subsection{Game Outcome and Post-Game Tremor}

To test whether game outcome (win/loss) and APM predicted post-game tremor, linear mixed models were fitted with each tremor indicator where ``won'' $\in \{0, 1\}$ indicates whether the player won or lost, and ``apm'' is the player's average actions per minute. Neither game outcome nor APM significantly predicted post-game tremor in any frequency band (all $p > .05$). A forearm-lateralization model including left/right forearm as a covariate was calculated, and yielded no significant effects.

\section{Discussion}
\label{sec:discussion}

Despite the clear limitations, our study showed that indeed, players in a natural tournament environment deviated from the reference values established with the same measurement protocol as used by other authors~\cite{Gajewski2023}. Nonetheless, the physiological tremor was not found to be tied to simple in-game indicators (victory/defeat). Interestingly, the post-game measurements, which might have indicated forearm fatigue over time, failed to show a significant relationship with the game outcome.

Esports players significantly deviated from reference values across all analyzed frequency bands: tremor power was reduced in physiologically relevant bands. In contrast, lower-frequency components showed an opposite trend. Similar reductions in variability or oscillatory activity have been reported in other expert populations that require fine motor skills, such as musicians and surgeons. Such adaptations were proven to be a result of training~\cite{Berger2025,Liu2024Shooting,Verrelli2016}. Tremor indicators also showed a gradual decrease over the course of the tournament day, particularly in higher frequency bands, a somewhat counterintuitive pattern, as fatigue is often associated with increased tremor amplitude in traditional motor tasks~\cite{Kulis2025Tremor}. Possible explanations include players' adaptation to the competitive environment over time, leading to improved motor stabilization or reduced variability; alternatively, sustained cognitive engagement and focus may suppress physiological tremor despite accumulating fatigue. This sustained reduction in tremor over time might also indicate enhanced neural efficiency in motor execution or effective stress mitigation strategies employed by professional esports athletes, distinguishing them from the general population~\cite{Gostilovich2023EsportsBiomarkers}.

Contrary to expectations, tremor measures were not significantly associated with match outcomes or APM. This surprising lack of detected relationship suggests a decoupling between raw physiological responses and overt performance metrics, potentially highlighting the compensatory role of cognitive and strategic elements in elite esports gameplay. Thus, physiological tremor may not be a direct predictor of short-term performance success in esports. The analysis did not reveal substantial differences in the rate of tremor changes between the left and right forearms. This bilateral symmetry in tremor response may indicate a generalized systemic reaction to competitive stress rather than localized neuromuscular fatigue, which typically manifests as asymmetrical changes~\cite{Park2022PhysTremor}.

A major strength of the present study is its high ecological validity, as tremor measurements were collected in a real tournament environment rather than in a controlled laboratory setting. This ``in-the-wild'' approach allows for capturing authentic psychophysiological responses during competitive gameplay, including the combined effects of cognitive load, stress, and motor demands.

\section{Limitations}
\label{sec:limitations}

Despite the nature of our study and the statistical significance of the findings, several limitations could not be addressed. First, due to the nature of the tournament environment, we were not able to maintain consistent timing between the start and end of the match and the tremor measurement. Second, because the players were mostly focused on their matches, some measurements might have been omitted. Third, not all players were measured an equal number of times, as this is naturally dependent on their tournament progression and performance. Fourth, the sample size of 16 players is relatively small. Nonetheless, we believe that our study showcases an interesting approach to measuring players ``in the wild'' to verify and uncover the true tournament-specific load, similar to ongoing research on amateur players in other esports titles \cite{Sadowska2023}.
\section{Conclusions}
\label{sec:conclusions}

Our study provides some insights into the physiological tremor characteristics of the esports players under tournament conditions. Keeping our initial goal and research questions in mind, we conclude that: \textbf{RQ1:} Firstly, we found that the characteristics of our esports player group were significantly different from the population norm. This effect can be explained as part of the adaptation process to the specific demands of esports, as seen in research on surgical training. While we initially expected the tremor to increase due to fatigue or the emotional load of the tournament, we ultimately observed a decreased tremor profile throughout the tournament. \textbf{RQ2:} Finally, we found no significant relationship between game outcome or APM and post-game tremor levels. This suggests that, at least at the level of individual game outcomes, the physiological tremor response does not differ between winning and losing, and that greater motor activity does not immediately translate into measurable changes in tremor after a game.

\subsection{Future Work}

Tremor analysis could offer a monitoring tool in esports, where physiological markers remain limited~\cite{Leis2020Psychological}. Future research should combine field measurements with controlled experiments to isolate the effects of fatigue, stress, and cognitive load on tremor characteristics. Furthermore, integrating additional physiological and psychological measures would provide a more holistic view of player load. Longitudinal studies tracking tremor patterns alongside training, experience, and skill development are also needed. Finally, exploring more sensitive performance metrics, such as reaction time and motor precision, may reveal closer links to tremor than overall win/loss outcomes and in-game performance.

\section*{Acknowledgements}

We wish to extend our gratitude towards Daniel ``danskyck'' Grabka and Adrian ``Kashim'' Główczyk  for his ongoing efforts in esports tournament organization, his collaboration, and allowing our presence during the tournaments that he has organized. Similarly, we wish to thank Marcin Warda, and Tomasz Kowalski who assisted in the process of infrastructure preparation and data collection, without their expertise this study would not be possible.
\section*{Declarations}

\subsection*{Authors' contributions}

\begin{itemize}
  \item Conceptualization: Andrzej Białecki, Jan Gajewski;
  \item Supervision: Andrzej Białecki, Jan Gajewski;
  \item Methodology: Andrzej Białecki, Jan Gajewski;
  \item Formal Analysis: Andrzej Białecki, Izabela Ghafour, Jan Gajewski;
  \item Investigation: Andrzej Białecki, Izabela Ghafour, Jan Gajewski;
  \item Writing - original draft: Andrzej Białecki, Izabela Ghafour, Szymon Kuliś, Maciej Skorulski;
  \item Writing - review and editing: Andrzej Białecki, Izabela Ghafour, Szymon Kuliś, Maciej Skorulski, Jan Gajewski;
  \item Data curation: Andrzej Białecki;
  \item Technical Oversight: Andrzej Białecki;
  \item Software: Andrzej Białecki, Izabela Ghafour;
\end{itemize}

\subsection*{Funding}
This publication was self-funded.

\subsection*{Conflicts of interest/Competing interests}
Authors declare no conflict of interest.

\bibliographystyle{IEEEtran}
\bibliography{IEEEabrv,sources.bib}

\end{document}